\documentclass{appolb}

\usepackage{calc}
\usepackage{graphicx}
\usepackage{dcolumn}
\usepackage{bm}
\usepackage{slashed}
\usepackage{amsmath,graphicx}
\usepackage[colorlinks=true,linktocpage=true,linkcolor=blue,citecolor=blue]{hyperref}
\usepackage{float}
\usepackage{nicefrac}
\usepackage[normalem]{ulem}
\usepackage{amsmath}
\usepackage{subfigure}
\usepackage{float} 
\usepackage{bbold}
\usepackage{cite} 

\def\Cv{{\boldsymbol C}}

\def\nU{n_{(0)}}
\def\eU{\varepsilon_{(0)}}
\def\PU{P_{(0)}}
\def\sU{s_{(0)}}

\newcommand{\sh}[1]{\sinh#1}
\newcommand{\ch}[1]{\cosh#1}

\def\HP{\hphantom{\alpha}} 

\newcommand{\beq}{\begin{eqnarray}}
\newcommand{\eeq}{\end{eqnarray}}

\newcommand{\bea}{\begin{eqnarray}}
\newcommand{\eea}{\end{eqnarray}}

\newcommand{\bel}[1]{\begin{eqnarray}\label{#1}}
\newcommand{\eel}{\end{eqnarray}}

\def\a{\alpha}
\def\b{\beta}
\def\g{\gamma}
\def\d{\delta} 

\def\s{\sigma}

\newcommand{\rf}[1]{Eq.~(\ref{#1})}
\newcommand{\rfm}[1]{Eqs.~(\ref{#1})}

\newcommand{\rfn}[1]{(\ref{#1})}

\newcommand{\nn}{\nonumber}

\newcommand{\p}{\partial}

\newcommand{\f}[2]{\frac{#1}{#2}}
\newcommand{\onehalf}{{\nicefrac{1}{2}}}

\renewcommand\sout{\bgroup \color{blue} \ULdepth=-.5ex \ULset}

\def\n0{n_{(0)}}
\def\e0{\varepsilon_{(0)}}
\def\P0{P_{(0)}}
\def\s0{s_{(0)}}

\def\LR{\left(} 
\def\RR{\right)}
\def\LS{\left[} 
\def\RS{\right]}

\def\HP{\hphantom{\alpha}} 

\def\fplusrsxp{f^+_{rs}(x,p)}

\def\fminusrsxp{f^-_{rs}(x,p)}

\def\bmu{\beta_\mu}

\def\omnL{\omega_{\mu\nu}}
\def\omnU{\omega^{\mu\nu}}

\def\pmu{p^\mu}

\def\pv{{\boldsymbol p}}

\def\SmunuU{{\Sigma}^{\mu\nu}}

\def\S0iU{{\Sigma}^{0i}} 
 
\def\SmnU{{\Sigma}^{\mu\nu}}

\def\CHI{\chi}

\def\ubarrp{{\bar u}_r(p)}

\def\usp{u_s(p)}
\def\urp{u_r(p)}

\def\vbarrp{{\bar v}_r(p)}
\def\vbarsp{{\bar v}_s(p)}
\def\vsp{v_s(p)}
\def\vrp{v_r(p)}

\def\g5{\gamma_5}

\def\Weqpmxk{{\cal W}^{\pm}_{\rm eq}(x,k)}

\def\Feqpmxk{{\cal F}^{\pm}_{\rm eq}(x,k)}

\def\Peqpmxk{{\cal P}^{\pm}_{\rm eq}(x,k)}

\begin{document} 
\title{
Relativistic hydrodynamics for spin-polarized media
\thanks{Presented by Radoslaw Ryblewski at XXVI Cracow Epiphany Conference, LHC Physics: Standard Model and Beyond, January 7-10, 2020, Krak\'ow, Poland.} 
}
\author{Radoslaw Ryblewski \and Rajeev Singh
		 \address{ 
		 	Institute of Nuclear Physics, PL-31342 Krak\'ow, Poland}
}
\maketitle
\begin{abstract} 
We summarize the key ingredients of the recently proposed formalism of relativistic perfect-fluid hydrodynamics with spin. Based on the underlying kinetic theory definitions for the equilibrium distribution functions we obtain the evolution equations governing the system's expansion. Employing Bjorken symmetry we study the spin polarization dynamics of the system.
\end{abstract}
%
%
\section{Introduction}
%
%
The first positive measurements of spin polarization of $\Lambda$ hyperons made lately by the STAR Collaboration \cite{STAR:2017ckg,Adam:2018ivw,Niida:2018hfw,Adam:2019srw} have revived the interest in the studies of the relation between the vorticity of the matter produced in relativistic heavy-ion collisions and the averange spin polarization of particles produced in these processes \cite{Becattini:2009wh,Becattini:2013fla,Montenegro:2017rbu,Montenegro:2017lvf,Becattini:2018duy,Boldizsar:2018akg,Prokhorov:2018bql,Yang:2018lew,Florkowski:2019voj,Weickgenannt:2019dks,Hattori:2019lfp,Ambrus:2019ayb,Sheng:2019kmk,Prokhorov:2019cik,Ivanov:2019wzg,Hattori:2019ahi,Xie:2019jun,Liu:2019krs,Wu:2019eyi,Becattini:2019ntv,Zhang:2019xya,Li:2019qkf,Florkowski:2019gio,Prokhorov:2019yft,Fukushima:2020qta,Liu:2020ymh,Bhadury:2020puc,Tabatabaee:2020efb,Liu:2020flb,Yang:2020hri,Deng:2020ygd,Taya:2020sej}; for a recent review see \cite{Becattini:2020ngo}. Recently, it has been shown that the thermal-based models~\cite{Becattini:2016gvu,Karpenko:2016jyx,Li:2017slc,Xie:2017upb} which correctly describe the global polarization unfortunately are not able to explain the differential observables \cite{Adam:2019srw}. These models are based on the assumption that the spin polarization of particles emitted at freeze-out is entirely determined by the quantity known as thermal vorticity~\cite{Becattini:2007sr,Becattini:2013fla} and do not include the possibility of its independent dynamical evolution, which may take place during the fluid expansion. In this work,  following ideas put forward in Refs.~\cite{Florkowski:2017ruc,Florkowski:2017dyn,Florkowski:2018myy,Florkowski:2018fap,Florkowski:2018ahw,Florkowski:2019qdp}, we study such possibility within the framework of relativistic perfect-fluid hydrodynamics with spin. 
%
%
\section{Equilibrium distribution functions}
%
%
Relativistic fluid dynamics may be derived from the underlying kinetic theory assuming that the distribution function describing the equilibrium state of the system is known \cite{Florkowski:2017olj}. Herein, following works by Becattini et al.  \cite{Becattini:2013fla},  we assume that the local equilibrium state of the relativistic system of particles ($+$) and antiparticles ($-$) with spin $\onehalf$ and mass $m$ is described by the following phase-space distribution functions (spin density matrices)
\bel{fplusrsxp}
\fplusrsxp =  \ubarrp X^+ \usp, \qquad
\fminusrsxp = - \vbarsp X^- \vrp,
\eel
where $x$ is the space-time position and $p$ is the four-momentum, and $\urp$ and $\vrp$ are Dirac bispinors ($r,s = 1,2$) with the normalization $\ubarrp \usp=\,\delta_{rs}$ and $\vbarrp \vsp=-\,\delta_{rs}$.

The matrices $X^{\pm}$ have the form of generalized relativistic Boltzmann distributions
\bel{XpmM}
X^{\pm} =  \exp\left[\pm \xi(x) - \bmu(x) \pmu  \pm \f{1}{2} \omnL(x)  \SmunuU \right], \nn
\eel
where $\beta^\mu \equiv U^\mu/T$ and $\xi \equiv \mu/T$, with $T$, $\mu$ and $U^\mu$ denoting the temperature, baryon chemical potential and four-velocity,  respectively. The quantity $\omnL$ is the spin polarization tensor satisfying $\omega^{\mu\nu}=-\omega^{\nu\mu}$ and $\SmunuU  \equiv  \f{i}{4} [\gamma^\mu,\gamma^\nu]$ is the spin operator.  

Employing expressions derived in Ref.~\cite{DeGroot:1980dk} and using definitions \rfn{fplusrsxp} we can determine the corresponding equilibrium Wigner functions
\bea
{\cal W}^\pm_{\rm eq}(x,k) &=& \frac{e^{\pm \xi}}{4 m}  \int dP
\,e^{-\beta \cdot p }\,\, \delta^{(4)}(k \mp p) \label{eq:Weqpxk2} \\
&& \times  \left[2m (m \pm \slashed{p}) \cosh(\zeta) \pm \f{\sinh(\zeta)}{2\zeta}  \, \omnL \,(\slashed{p} \pm m) \SmnU (\slashed{p} \pm m) \right], \nn
\eea
where $k$ is the off-mass-shell four-momentum of particles, 
$dP = d^3p/((2 \pi )^3 E_p)$  with $E_p = \sqrt{m^2 + \pv^2}$ being the on-shell particle energy, and $\zeta =  \f{1}{2 \sqrt{2}} \sqrt{ \omnL \omnU}$. 

It is convenient to consider the Clifford-algebra expansion of the Wigner function \rfn{eq:Weqpxk2} 
\bea
\Weqpmxk &=& \f{1}{4} \left[ \Feqpmxk + i \gamma_5 \Peqpmxk + \gamma^\mu {\cal V}^\pm_{{\rm eq}, \mu}(x,k) \right. \nn \\
&& \left.  \hspace{1cm} + \gamma_5 \gamma^\mu {\cal A}^\pm_{{\rm eq}, \mu}(x,k)
+ \SmnU {\cal S}^\pm_{{\rm eq}, \mu \nu}(x,k) \right], \nn
\label{eq:wig_expansion}
\eea
where the coefficient functions $\mathcal{X} \in\left\{\mathcal{F}, \mathcal{P}, \mathcal{V}_{\mu}, \mathcal{A}_{\mu}, \mathcal{S}_{\nu \mu}\right\}$ can be extracted by calculating the trace of $\Weqpmxk$ multiplied first by: $\left\{\mathbf{1},-i \gamma_{5}, \gamma_{\mu}, \gamma_{\mu} \gamma_{5}, 2 \Sigma_{\mu \nu}\right\}$.
%
%
\section{Kinetic equations}
%
%
General Wigner function satisfies the kinetic equation 
\bel{eq:eqforWC}
\left(\gamma_\mu K^\mu - m \right) {\cal W}(x,k) = C[{\cal W}(x,k)],
\eel
where the differential operator reads $K^\mu = k^\mu + \frac{i \hbar}{2} \,\p^\mu$. 
In the case of global equilibrium, the Wigner function satisfies exactly \rf{eq:eqforWC} with $C[{\cal W}(x,k)] = 0$. The usual treatment of \rf{eq:eqforWC}  is to consider the semi-classical expansion of the coefficient functions 
\bel{eq:semi}
\mathcal{X}=\mathcal{X}^{(0)}+\hbar \mathcal{X}^{(1)}+\hbar^{2} \mathcal{X}^{(2)}+\cdots. \nn
\eel
The analysis of \rf{eq:eqforWC} up to the next-to-leading order in $\hbar$ yields the following kinetic equations for the two independent coefficients $\mathcal{F}_{\mathrm{eq}}$ and $\mathcal{A}_{\mathrm{eq}}^{\nu}$, 
\bel{eq:ke}
k^{\mu} \partial_{\mu} \mathcal{F}_{\mathrm{eq}}(x, k)=0, \quad k^{\mu} \partial_{\mu} \mathcal{A}_{\mathrm{eq}}^{\nu}(x, k)=0, \quad k_{\nu} \mathcal{A}_{\mathrm{eq}}^{\nu}(x, k)=0.
\eel
In global equilibrium  \rfm{eq:ke} are exactly fulfilled which results in the conditions that $\beta_{\mu}$ is a Killing vector, and $\xi$ and $\omega_{\mu \nu}$ are constant, however $\omega_{\mu \nu}$ does not have to be equal to  thermal vorticity $\varpi_{\mu \nu} = -\frac{1}{2} \left(\p_\mu \beta_\nu - \p_\nu \beta_\mu \right) = \hbox{const}$.
%
%
\section{Hydrodynamic equations}
%
%
In local equilibrium \rfm{eq:ke} are not satisfied exactly. In this case, we 
adopt the standard treatment \cite{Denicol:2012cn}, namely,  by allowing for $x$ dependence of the $\beta$, $\xi$ and $\omega$, we require that only certain moments in momentum space of the kinetic equations \rfn{eq:ke} are satisfied. This method leads to equations expressing conservation laws for charge, energy, linear momentum and spin\cite{Florkowski:2018ahw}
\begin{eqnarray} 
\quad\partial_\mu N^\mu = 0, \label{Ncons} \\
\partial_\mu T^{\mu\nu}_{\rm GLW} = 0, \label{Tcons}\\
\partial_\lambda  S_{\rm GLW}^{\lambda, \alpha \beta} =0,
\label{Scons}
\end{eqnarray}
where the baryon current, the energy-momentum tensor, and the spin tensor are given by the de Groot - van Leeuwen - van Weert (GLW)~\cite{DeGroot:1980dk}  expressions
\bea
N^\alpha &=& n U^\alpha,\\
 T^{\a\b}_{\rm GLW} &=& (\varepsilon + P ) U^\a U^\b - P g^{\a\b},\\
 S^{\alpha , \beta \gamma }_{\rm GLW}
&=& \cosh(\xi) \LS n_{(0)} U^\alpha \omega^{\beta\gamma}  +  {\cal A}_{(0)} \, U^\a U^\d U^{[\b} \omega^{\gamma]}_{\HP\d}  \right. \\
&& +\left. \, {\cal B}_{(0)} \, \Big( 
U^{[\b} \Delta^{\a\d} \omega^{\gamma]}_{\HP\d}
+ U^\a \Delta^{\d[\b} \omega^{\gamma]}_{\HP\d}
+ U^\d \Delta^{\a[\b} \omega^{\gamma]}_{\HP\d}\Big) \RS,
\eel
where $\Delta^{\mu\nu} =g^{\mu\nu} - U^\mu U^\nu$ is the projector on the space orthogonal to $U$.

In the leading order in the polarization tensor the energy density $\varepsilon$, the pressure $P$, and the baryon density $n$ are given by the formulas
\bea 
n &=& 4 \, \sinh(\xi)\, \nU(T), \label{n0small} \\
\varepsilon &=& 4 \, \cosh(\xi) \, \eU(T),\label{e0small}\\ 
P &=& 4 \, \cosh(\xi) \, \PU(T), \label{P0small} 
\eea
where we defined the auxiliary quantities describing thermodynamic properties of the system of spin-less and neutral massive Boltzmann particles\cite{Florkowski:2010zz}
\beq
\nU(T) &=&  \f{ T^3}{2\pi^2}\,  \hat{m}^2 K_2\left( \hat{m}\right), \label{n0c}\\
\eU(T) &=& \f{ T^4 }{2\pi^2}  \, \hat{m} ^2
 \Big[ 3 K_{2}\left( \hat{m} \right) + \hat{m}  K_{1} \left( \hat{m}  \right) \Big],  \label{e0c}\\
\PU(T) &=& T \, \nU(T) . \label{P0c}
\eeq
 The quantities ${\cal B}_{(0)} $ and ${\cal A}_{(0)}$ are defined as follows 
\beq
{\cal B}_{(0)} =-\frac{2}{\hat{m}^2} s_{(0)}(T) ,\qquad
{\cal A}_{(0)}  = -3{\cal B}_{(0)} +2 n_{(0)}(T) 
\eeq
with $\sU =   \LR\eU+\PU\RR / T$ being the entropy density and $\hat{m}=m/T$. 
%
%
\section{Bjorken expansion}
%
%
Similarly to the Faraday tensor the polarization tensor $\omega_{\mu\nu}$ may be decomposed into electric-like ($\kappa$) and magnetic-like ($\omega$) components
\beq
\omega_{\mu\nu} &=& \kappa_\mu U_\nu - \kappa_\nu U_\mu + \epsilon_{\mu\nu\a\b} U^\a \omega^{\b},\label{spinpol1}
\eeq
where the four-vectors $\kappa$ and $\omega$ satisfy the conditions 
\beq
\kappa\cdot U = 0,\qquad \omega \cdot U = 0.
\label{orth}
\eeq
In the case of transversely homogeneous systems undergoing boost-invariant expansion in the longitudinal direction, also known as the Bjorken flow \cite{Bjorken:1982qr}, it is convenient to introduce the following four-vector basis 
\beq
U^\a &=& \frac{1}{\tau}\LR t,0,0,z \RR = \LR \cosh(\eta), 0,0, \sinh(\eta) \RR, \nn \\
X^\a &=& \LR 0, 1,0, 0 \RR,\nn\\
Y^\a &=& \LR 0, 0,1, 0 \RR, \nn\\
Z^\a &=& \frac{1}{\tau}\LR z,0,0,t \RR = \LR \sinh(\eta), 0,0, \cosh(\eta) \RR, 
\label{BIbasis}
\eeq
where $\tau = \sqrt{t^2-z^2}$ is the longitudinal proper time and $\eta = \half \ln((t+z)/(t-z))$ is the space-time rapidity.

The basis \rfn{BIbasis} satisfies the conditions
\begin{eqnarray}
 &&U \cdot U = 1,\nn\\
X \cdot X &=&   Y \cdot Y \,\,=\,\, Z \cdot Z \,\,=\,\, -1,\nn \\ \label{orthXYZ}
X \cdot U  \,\,&=& Y \cdot U\,\, \,\,=\,\, Z \cdot U \,\,=\,\, 0,   \\
X \cdot Y  &=&  Y \cdot Z \,\,=\,\, Z \cdot X \,\,=\,\, 0.  \nn
\end{eqnarray}
Using \rfm{orth} and \rfm{orthXYZ}, one can decompose the vectors $\kappa^{\mu}$ and $\omega^{\mu}$ as follows
\beq
\kappa^\a &=&  C_{\kappa X} X^\a + C_{\kappa Y} Y^\a + C_{\kappa Z} Z^\a,  \nn\\
\omega^\a &=&  C_{\omega X} X^\a + C_{\omega Y} Y^\a + C_{\omega Z} Z^\a, \label{decom}
\eeq
where the coefficients ${C}_{\kappa X}$, ${C}_{\kappa Y}$, ${C}_{\kappa Z}$, ${C}_{\omega X}$, ${C}_{\omega Y}$, and ${C}_{\omega Z}$ are scalar functions of proper time solely.  

Using \rfm{decom} in \rf{Scons} and projecting the latter on $U_\mu X_\nu$, $U_\mu Y_\nu$, $U_\mu Z_\nu$, $X_\mu Y_\nu$, $X_\mu Z_\nu$ and $Y_\mu Z_\nu$, we obtain the following six evolution equations
\begin{equation}
{\rm diag}\LR
\cal{L}, \cal{L}, \cal{L}, \cal{P}, \cal{P}, \cal{P}\RR \,\,
\Dot{\Cv} ={\rm diag}\LR
{\cal{Q}}_1, {\cal{Q}}_1, {\cal{Q}}_2, {\cal{R}}_1, {\cal{R}}_1, {\cal{R}}_2 \RR\,\,
\Cv, \label{cs}
\end{equation} 
where $\Cv = \LR C_{\kappa X}, C_{\kappa Y}, C_{\kappa Z},  C_{\omega X}, C_{\omega Y}, C_{\omega Z} \RR$, $\dot{(\dots)} \equiv U \cdot \p = \p_\tau$ and  
\beq
{\cal L}(\tau)&=&{\cal A}_1-\frac{1}{2}{\cal A}_2-{\cal A}_3,\nn\\
{\cal P}(\tau)&=&{\cal A}_1,\nn\\
{\cal{Q}}_1(\tau)&=&-\left[\dot{{\cal L}}+\frac{1}{\tau}\left( {\cal L}+ \frac{1}{2}{\cal A}_3\right)\right],\nn\\
 {\cal{Q}}_2(\tau)&=&-\left(\dot{{\cal L}}+\frac{{\cal L}}{\tau}   \right),\nn\\
  {\cal{R}}_1(\tau)&=&-\left[\Dot{\cal P}+\frac{1}{\tau}\left({\cal P} -\frac{1}{2} {\cal A}_3 \right) \right],\nn\\
 {\cal{R}}_2(\tau)&=&-\left(\Dot{{\cal P}} +\frac{{\cal P}}{\tau}\right).\nn
 \label{LPQR}
 \eeq
 with 
 \beq
{\cal A}_1 &=& \cosh(\xi) \LR \nU -  {\cal B}_{(0)} \RR \label{A1} ,\nn\\ 
{\cal A}_2 &=& \cosh(\xi) \LR {\cal A}_{(0)} - 3{\cal B}_{(0)} \RR \nn \label{A2} , \\ 
{\cal A}_3 &=& \cosh(\xi)\, {\cal B}_{(0)}\label{A3},\nn
\eeq

From \rfm{cs} we observe that in the case of Bjorken expansion the ${C}$ coefficients evolve independently. Due to the rotational symmetry in the transverse plane the functions ${C}_{\kappa X}$ and ${C}_{\kappa Y}$ (as well as ${C}_{\omega X}$ and ${C}_{\omega Y}$) obey the same differential equations.

Employing the Bjorken symmetry,  the conservation of the charge current \rfn{Ncons} can be written as
\beq
\frac{dn}{d\tau}+\frac{n}{\tau}=0\label{ns}
\eeq
while the conservation of energy and linear momentum \rfn{Tcons}  (projected on $U$) gives
\beq
\frac{d\varepsilon}{d\tau}+\frac{(\varepsilon+P)}{\tau}=0.\label{ts}
\eeq 
%
\section{Spin polarization of particles at freeze-out}
%
%
The information about the space-time evolution of the spin polarization tensor may be used to determine the average spin polarization per particle which is defined as follows~\cite{Florkowski:2018ahw}
\beq
\langle\pi_{\mu}\rangle=E_p\frac{d\Pi _{\mu }(p)}{d^3 p}/E_p\frac{d{\cal{N}}(p)}{d^3 p}, \label{pi}
\eeq
where $E_p\frac{d\Pi _{\mu }(p)}{d^3 p}$ is the total value (integrated over freeze out hypersuface $\Delta \Sigma_{\lambda }$) of the Pauli-Luba\'nski vector,
\beq
E_p\frac{d\Pi _{\mu }(p)}{d^3 p} = -\f{ \cosh(\xi)}{(2 \pi )^3 m}
\int
\Delta \Sigma _{\lambda } p^{\lambda } \,
e^{-\beta \cdot p} \,
\tilde{\omega }_{\mu \beta }p^{\beta },  \nn
\eeq
and 
\beq
E_p\frac{d{\cal{N}}(p)}{d^3 p}&=&
\f{4 \cosh(\xi)}{(2 \pi )^3}
\int
\Delta \Sigma _{\lambda } p^{\lambda } 
\,
e^{-\beta \cdot p}, \nn
\eeq
is the momentum density of particles and antiparticles.

The polarization vector $\langle\pi^{\star}_{\mu}\rangle$ in the particle rest rame is obtained by performing the canonical boost \cite{Leader:2001gr}  of \rfn{pi}
\beq
\langle\pi^{\star}_{\mu}\rangle=-\frac{1}{8m }\left[\begin{array}{c}
0 \\ \\
\left(\frac{p_x\sh y_p }{b}\right) a_1 +
  \LR\frac{ \chi \,p_x \ch y_p  }{b} \RR a_2\!+\!2 C_{\kappa Z} p_y  \!-\!\chi C_{\omega X}{m}_T \\ \\
\left(\frac{p_y\sh y_p }{b}\right)a_1+   \LR\frac{ \chi \,p_y \ch y_p  }{b} \RR a_2\!-\!2 C_{\kappa Z} p_x \!-\!\chi C_{\omega Y}{m}_T \\ \\
 -\left(\frac{m\ch y_p+m_T}{b}\right)a_1
-\LR\frac{\chi \,m\,\sh y_p}{b}\RR a_2 \nn\\
\end{array}
\right],\\
\label{PLVPPLPRF}  
\eeq 
where $a_1=\chi\left(C_{\kappa X} p_y-C_{\kappa Y} p_x\right)+2 C_{\omega Z} m_T$, $a_2=C_{\omega X} p_x+C_{\omega Y} p_y$, $b=m_T \ch y_p+m$, and $\CHI=\left( K_{0}\left( \hat{m}_T \right)+K_{2}\left( \hat{m}_T \right)\right)/K_{1}\left( \hat{m}_T \right)$ with  $\hat{m}_T=m_T/T$.  Above we used the following parametrization of the four-momentum  $p^\lambda = \left( m_T\ch y_p,p_x,p_y,m_T\sh y_p \right)$.

\begin{figure}
\centering
 \includegraphics[width=0.6\textwidth]{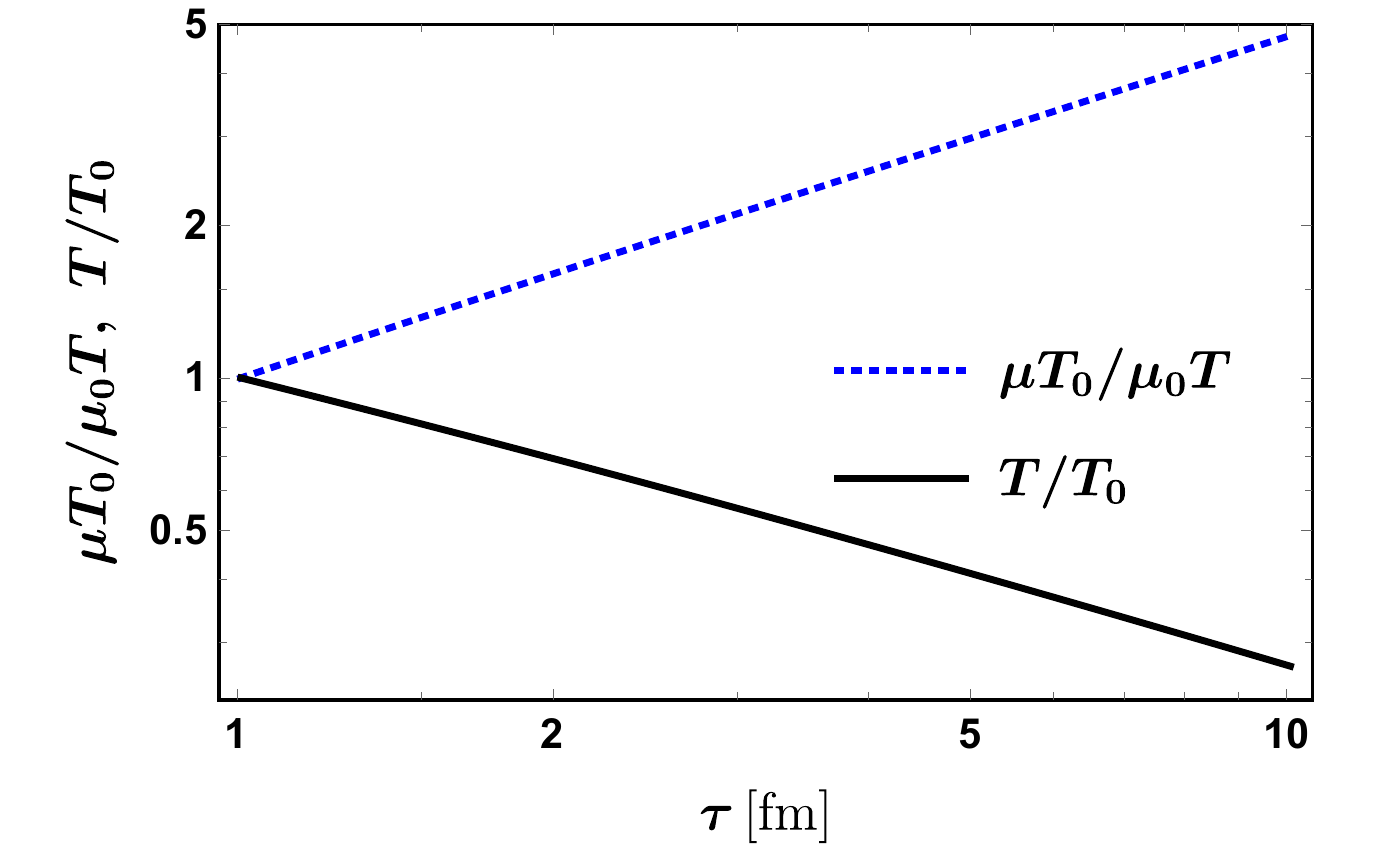} 
\caption{Proper-time dependence of the temperature scaled by its initial value (solid black line) and the ratio of baryon chemical potential over temperature rescaled by the initial ratio (dotted blue line).}
\label{fig:Tmu}
\end{figure}

\begin{figure}
\centering
 \includegraphics[width=0.6\textwidth]{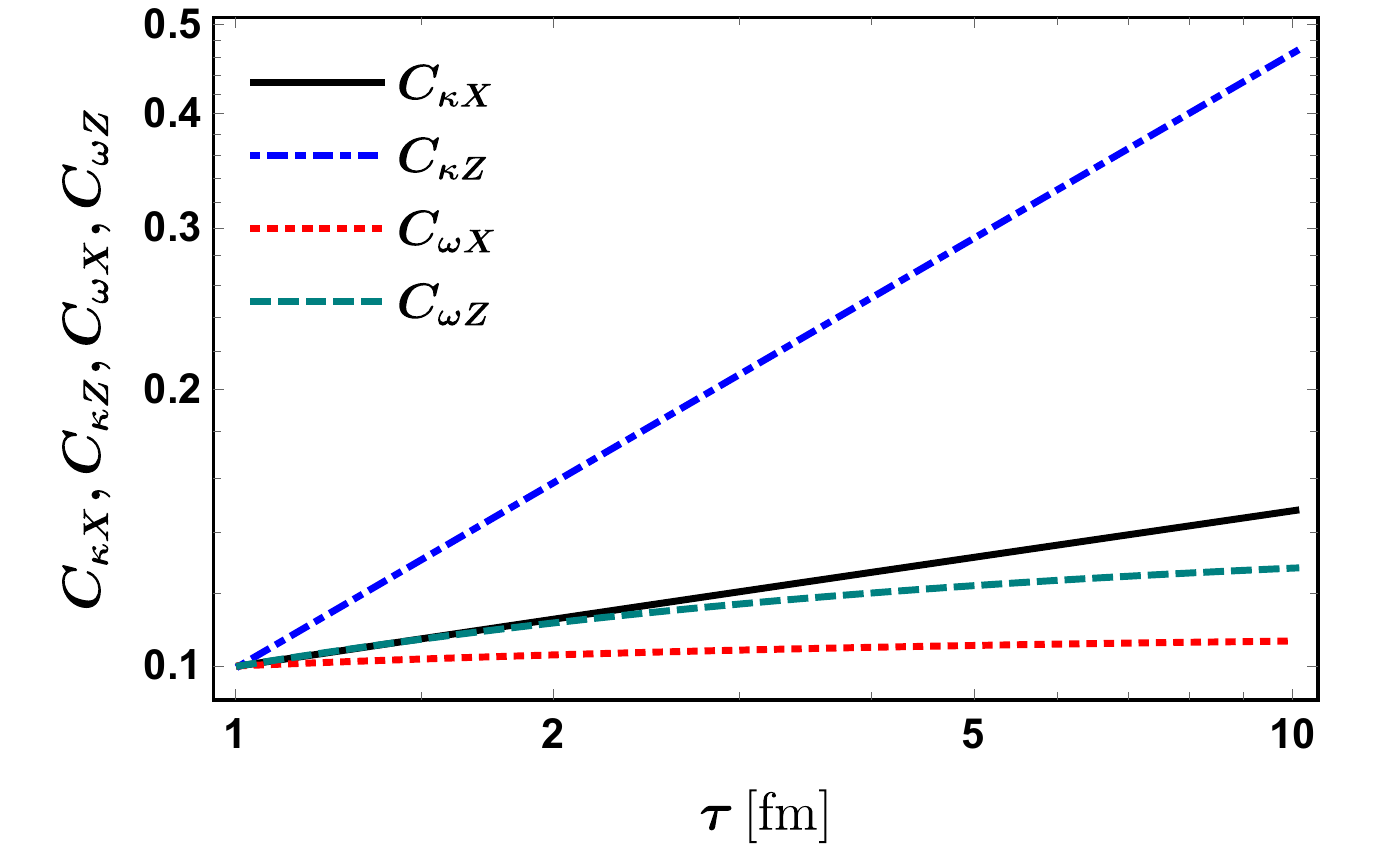} 
\caption{Proper-time dependence of the coefficients $C_{\kappa X}$ (solid black line), $C_{\kappa Z}$ (dashed-dotted blue line), $C_{\omega X}$ (dotted red line) and $C_{\omega Z}$ (dashed green line).}
\label{fig:C}
\end{figure}
%
\section{Results}
%
%
In this section we present the results obtained by solving numerically the differential equations \rfn{cs}, \rfn{ns}, and \rfn{ts}. We initialize the system at the proper time $\tau_0 = 1$ fm with initial temperature $T_0=T(\tau_0)=150$ MeV and the baryon chemical potential $\mu_0=\mu(\tau_0)=800$ MeV. We assume that the system consists of $\Lambda$ particles with mass $m=1116$ MeV. In Fig.~\ref{fig:Tmu}, we show the proper-time dependence of the (properly scaled) temperature and baryon chemical. We reproduce the well known results that the temperature of such a system decreases with proper-time while the ratio of chemical potential and temperature increases. In Fig.~\ref{fig:C}, we show the proper time dependence of the $C$ coefficients that describe the evolution of the spin polarization.

\begin{figure}
\centering
 \includegraphics[width=0.32\textwidth]{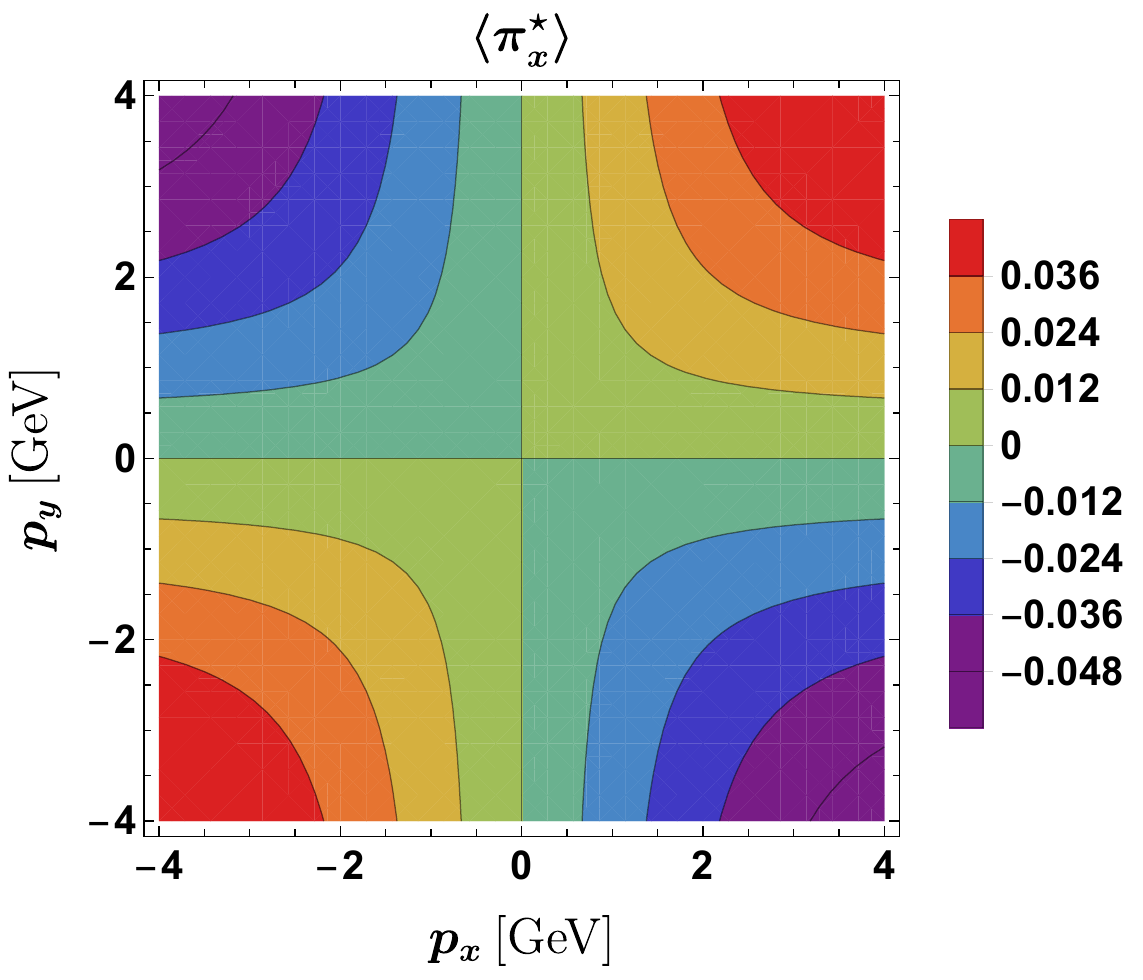}
\includegraphics[width=0.32\textwidth]{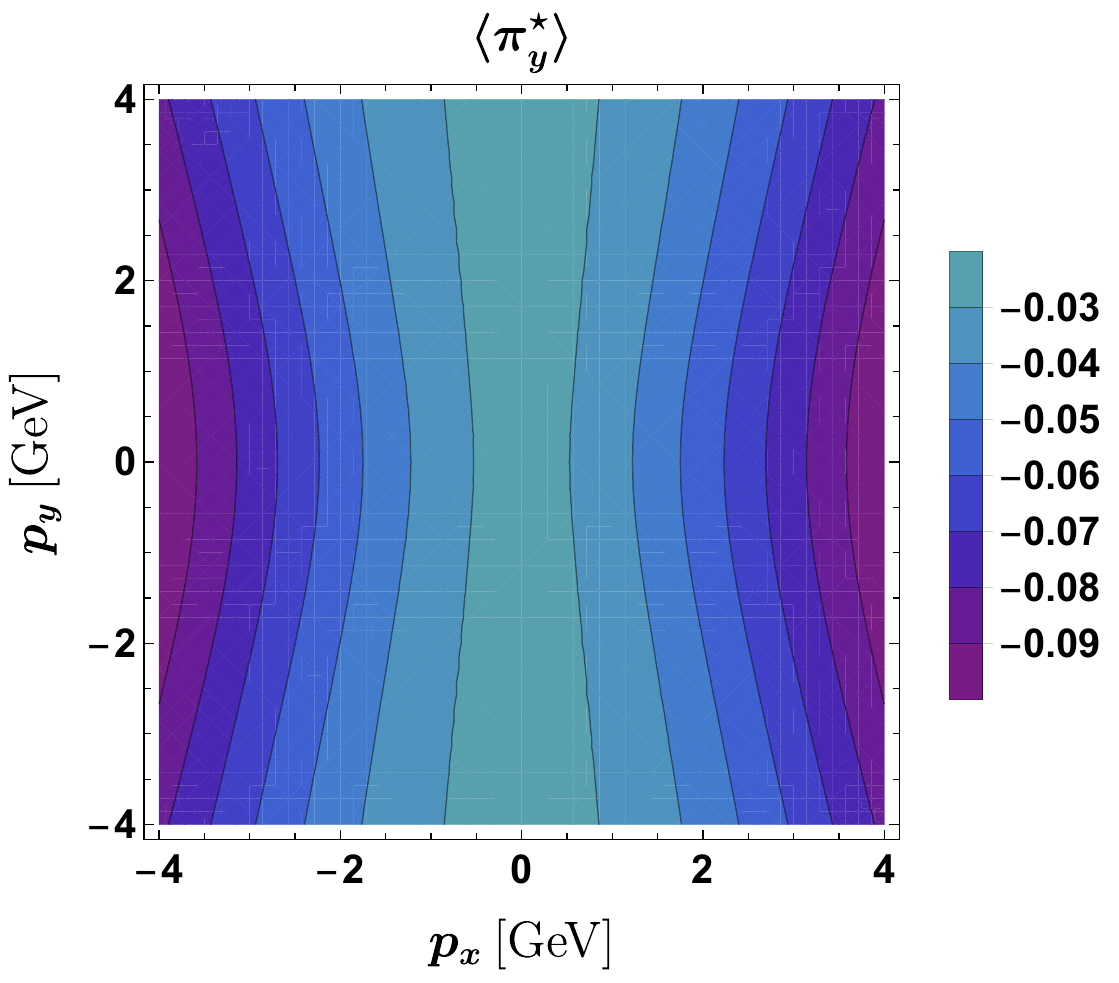}
 \includegraphics[width=0.32\textwidth]{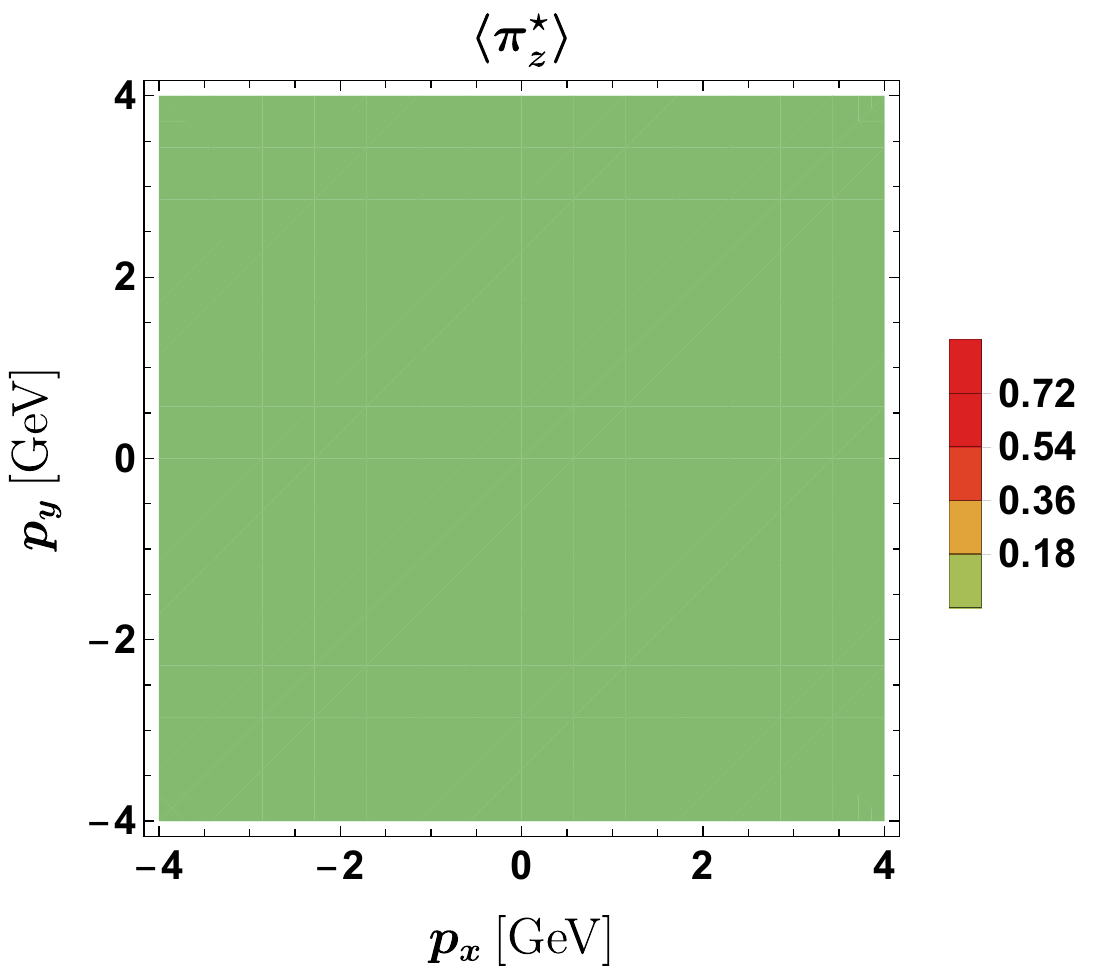}
\caption{Components of the particle-rest-frame mean polarization three-vector of $\Lambda$ particles obtained with the initial conditions $\mu_0=800$~MeV,
$T_0=155$~MeV, $C_{\kappa, 0}=(0,0,0)$ and $ C_{\omega, 0}=(0,0.1,0)$ for $y_p=0$.}
\label{fig:polarization1}
\end{figure}

The knowledge of the evolution of thermodynamic parameters and $C$ coefficients allows us to calculate the components of the particle-rest-frame mean polarization vector $\langle\pi^{\star}_{\mu}\rangle$ at freeze-out as functions of particle three-momentum, see Fig.~\ref{fig:polarization1}.
We observe that the component $\langle\pi^{\star}_{y}\rangle$ is negative, which reflects the initial spin polarization of the system.  Due to the Bjorken symmetry the longitudinal component ($\langle\pi^{\star}_{z}\rangle$) is vanishing which does not agree with the characteristic quadrupole structure of the longitudinal polarization observed in the experiment. One the other hand we observe $\langle\pi^{\star}_{x}\rangle$ exhibits quadrupole structure. Clearly, we observe that the Bjorken symmetry is too restrictive to address the experimental measurements correctly.

%
%
\section{Summary}
%
%
In this work we briefly reviewed the basic  ingredients  of the recently formulated  approach of relativistic perfect-fluid hydrodynamics with spin. Using the kinetic theory definitions for the local equilibrium distribution functions we derived  the evolution equations governing the system's expansion. Assuming the Bjorken flow of the matter we studied numerically the spin polarization dynamics of the system.
We have shown that the coefficient functions characterizing the spin polarization evolve independently. We have used these results to determine the spin polarization of particles at the freeze-out. We have shown that  within the simple Bjorken setup the characteristic features observed in the experiment can not be properly reproduced.
%
\section*{Acknowledgments}
%
Supported in part by the Polish National Science Center Grants No.   2016/23/B/ST2/00717 and No. 2018/30/E/ST2/00432.  
%

\end{document}